\documentclass[bibyear]{aa} % if the references are not structured 
\usepackage{graphicx}
\usepackage{txfonts}
\begin{document}

   \title{Evidence of a truncated spectrum in the angular correlation \\ function of the cosmic microwave background}

   \titlerunning{Angular Correlation of the CMB}

   \subtitle{}

   \author{F. Melia\inst{1} \and M. L\'opez-Corredoira\inst{2,3}}

   \institute{$^1$ Department of Physics, The Applied Math Program, and Department of Astronomy,
The University of Arizona, AZ 85721, USA\\
$^2$ Instituto de Astrof\'\i sica de Canarias,
E-38205 La Laguna, Tenerife, Spain\\
$^3$ Departamento de Astrof\'\i sica, Universidad de La Laguna,
E-38206 La Laguna, Tenerife, Spain}

   \date{Accepted December 21, 2017}

% \abstract{}{}{}{}{} 
% 5 {} token are mandatory
 
  \abstract
   {}
   {The lack of large-angle correlations in the fluctuations of the cosmic
microwave background (CMB) conflicts with predictions of slow-roll inflation.
But though probabilities ($\lesssim 0.24\%$) for the missing correlations
disfavor the conventional picture at $\gtrsim 3\sigma$, factors not
associated with the model itself may be contributing to the tension.
Here we aim to show that the absence of large-angle correlations is best 
explained with the introduction of a non-zero minimum wavenumber 
$k_{\rm min}$ for the fluctuation power spectrum $P(k)$.}
   {We assume that quantum fluctuations were generated in the early Universe 
with a well-defined power spectrum $P(k)$, though with a cutoff $k_{\rm min}\not=0$. 
We then re-calculate the angular correlation function of the CMB
and compare it with Planck observations.}
   {The Planck 2013 data rule out a zero
$k_{\rm min}$ at a confidence level exceeding $8\sigma$. Whereas purely
slow-roll inflation would have stretched all fluctuations beyond the
horizon, producing a $P(k)$ with $k_{\rm min}=0$---and therefore strong
correlations at all angles---a $k_{\rm min}\not=0$ would signal the presence
of a maximum wavelength at the time ($t_{\rm dec}$) of decoupling. This
argues against the basic inflationary paradigm---perhaps even suggesting
non-inflationary alternatives---for the origin and growth of perturbations
in the early Universe. In at least one competing cosmology, the $R_{\rm h}=ct$
universe, the inferred $k_{\rm min}$ corresponds to the gravitational
radius at $t_{\rm dec}$.}
   {}


   \maketitle
%
%________________________________________________________________

\section{Introduction}
The {\it Wilkinson} Microwave Anisotropy Probe (WMAP) (Bennett et al. 2003; Spergel
et al. 2003) and {\it Planck} (Planck 2014a), have resoundingly
confirmed the existence of several anomalies seen on very large scales by the
Cosmic Background Explorer (COBE) (Wright et al. 1996; Hinshaw et al. 1996). The 
most prominent among these is the lack of any significant correlation measured 
at angles $\gtrsim 60^\circ$. Its possible impact on the basic inflationary paradigm 
(Guth 1981; Linde 1982) has spurred a prolonged debate concerning whether it is due 
to a real physical phenomenon or unrecognized observational systematic effects.

The absence of large-angle correlations may simply be due to cosmic variance 
(Bennett et al. 2013; Copi et al. 2015), though probabilities for the missing 
correlations are typically $\lesssim 0.24\%$, disfavoring the basic inflationary 
picture at better than $3\sigma$ (see also Kim \& Naselsky 2011; Melia 2014; Gruppuso 
et al. 2016). This anomaly may also be due to subtleties in the foreground subtraction, 
or unrecognized instrumental systematics (Bennett et al. 2013), but this is far from 
settled.

The absence of large-angle correlations in the high-fidelity CMB maps poses
one of the most serious challenges to the basic inflationary paradigm and,
with it, to the internal self-consistency of the standard model. Since no
theoretical corrections can improve the fit (Planck 2014a), 
the largest angular scales are probing different physics than the anisotropies 
seen at angles smaller than $\sim 2^\circ$ which, in contrast, are highly 
consistent with the predictions of the standard model.

In parallel with this dichotomy between the small- and large-angle
correlations, WMAP and {\it Planck} have also revealed an unexpectedly weak
power in the low-$\ell$ multipoles (see Eq.~6 below) compared with the
corresponding power at higher $\ell$'s (see, e.g., Bennett et al. 2011).
This anomaly may or may not be related to the absence of angular correlation
at angles $\ge 60^\circ$ (Copi et al. 2007, 2015). Arguments have been made
on both sides though, if unrelated, the existence of two such anomalies 
significantly exacerbates the tension with the predictions of standard 
inflationary cosmology (see also Efstathiou 2004). The power deficit at 
large angular scales manifests itself in several ways, however, not just
via these two particular facets, so its impact on the interpretation
of the CMB fluctuations is extensive. A thorough consideration of the 
broader issues associated with the low angular power may be found in 
several recent publications by the Planck Collaboration (2014b, 2016).
In this paper, our focus will be specifically on the interpretation 
of the angular correlation function, which may now be studied at an
unprecedented level of precision.

Previous attempts 
at modifying the basic inflationary paradigm to address this issue have relied on the 
inflaton field evolving through an early fast-rolling stage, producing a characteristic 
scale when the wavenumber associated with the transition from fast to slow roll exited 
the horizon (Contaldi et al. 2003; Destri et al. 2010; Gruppuso et al. 2016). 
However, given the relative imprecision of the data avalaible then (compared to the
exquiste measurements provided most recently by {\it Planck}), the existence
of such a scale could be established at no more than $\sim 2\sigma$. 

In one such attempt to determine whether the power spectrum is truncated,
Niarchou et al. (2004) adopted a functional form (first introduced by Contaldi 
et al. 2003) 
\begin{equation}
P(k) = P_0(k)[1-e^{-(k/k_c)^\alpha}]\;,
\end{equation}
where 
\begin{equation}
P_0(k)=Ak^n
\end{equation}
is the usual primordial power-law spectrum, $k_c$ is a characteristic wavenumber, 
and $\alpha=1.8$, and carried out a Bayesian model comparison based on the
CMB power spectrum itself (rather than the angular correlation function) to 
demonstrate that the WMAP data available at that time preferred an attenuated $P(k)$ with 
$k_c\approx(5-6)\times 10^{-4}$ Mpc$^{-1}$. If the last scattering surface occurred at
$z_{\rm cmb}\sim 1080$ (according to the standard model), for which the expansion factor 
in a flat Universe was $a(z_{\rm cmb})= 1/(1+z_{\rm cmb})\approx 9.25\times 10^{-4}$, this
characteristic wavenumber would have corresponded to a physical fluctuation size 
$\lambda_{\rm max}\sim 10$ Mpc. But for reasons we shall describe shortly, this use of
the entire spectrum does not emphasize the large-angle anomalies, and given the somewhat
lower precision of WMAP (compared to {\it Planck}), Niarchou et al. (2004) concluded that 
a cutoff model such as Equation~(1) is preferred over one without a $k_c$ at only the 
$\sim 1.4\sigma$ level.

In this paper, we address the observed lack of angular correlation at large angles
with a clear, unobstructed focus on the possible existence of a cutoff
in the fluctuation spectrum as seen primarily at angles $\gg 1^\circ$, 
bolstered by the unprecedented accuracy of the 
{\it Planck} 2013 measurements. This distinctly different approach to the
resolution of this anomaly avoids an unnecessary reliance on inflation, which may
or may not have actually happened. A clear emergence of a non-zero $k_{\rm min}$ in
the {\it Planck} data would motivate the search for new physics in both inflationary
and non-inflationary scenarios. We note that the horizon problem plaguing $\Lambda$CDM
manifests itself only in models with an early period of deceleration, so inflation
is not required in all Friedmann-Robertson-Walker cosmologies. For example, it is
not present in models, such as the $R_{\rm h}=ct$ universe (Melia 2007, 2016, 2017;
Melia \& Shevchuk 2012), in which opposite sides of the sky reached thermal equilibrium 
following the big bang (Melia 2013). In this paper, we seek to uncover compelling
evidence in favor of such new physics beyond conventional, slow-roll inflation.

\section{Theoretical Background}
To implement the introduction of a $k_{\rm min}$, we assume that quantum
fluctuations were generated in the early Universe with a well-defined power spectrum
$P(k)$, and that these seeds subsequently grew linearly towards $t_{\rm dec}$. But
unlike the conventional picture, we will find that the most
important property of $P(k)$ that alleviates the anomaly is a non-zero value of the
minimum wavenumber $k_{\rm min}$, possibly due to an early transition from fast to slow
roll, or generic to a variety of non-inflationary scenarios. The $R_{\rm h}=ct$ universe,
which has been studied extensively in recent years, meets these criteria, so we know of
at least one alternative cosmology with the necessary characteristics (Melia 2016).

In the current $\Lambda$CDM, the fluctuations would have grown to all observable scales
as a result of the rapid expansion during inflation, so $k_{\rm min}=0$. In contrast,
a non-inflationary expansion would have had $k_{\rm min}\not=0$ if the fluctuation
growth was restricted to a finite range of wavelengths. Again resorting to
$R_{\rm h}=ct$ as an example, $k_{\rm min}$ could have corresponded to $2\pi$ times
the Hubble radius $R_{\rm h}(t_{\rm dec})$ at $t_{\rm dec}$. A
similar correspondence would have applied to the formation of structure from
topological defects (Brandenberger 1994). The results of our analysis should be quite
general and relevant to any modified mechanism of inflation, or to any non-inflationary
cosmology that possesses a non-zero minimum wavenumber in its power spectrum $P(k)$.

We follow convention and assume that a Gaussian random field in the plane of the sky
describes the microwave temperature $T(\hat{\bf n})$ in every direction
$\hat{\bf n}$, and write it as an expansion in spherical harmonics $Y_{lm}(\hat{\bf n})$.
The corresponding random coefficients $a_{lm}$ have zero mean, so the angular
correlation function linking directions $\hat{\bf n}_1$ and $\hat{\bf n}_2$ depends
only on $\cos\theta\equiv \hat{\bf n}_1\cdot \hat{\bf n}_2$, which we expand in
terms of Legendre polynomials:
\begin{equation}
C(\cos\theta)\equiv \langle T(\hat{\bf n}_1)T(\hat{\bf n}_2)\rangle=
\frac{1}{4\pi}\sum_\ell (2\ell+1)C_\ell P_\ell(\cos\theta)\;.
\end{equation}
The coefficients are statistically independent, so
\begin{equation}
\langle a^*_{\ell m}a_{\ell'm'}\rangle\propto\delta_{\ell\ell'}\,\delta_{mm'}\;,
\end{equation}
and statistical isotropy guarantees that the constant of proportionality
depends only on $\ell$:
\begin{equation}
\langle a^*_{\ell m}a_{\ell'm'}\rangle=C_\ell\,\delta_{\ell\ell'}\,\delta_{mm'}\;.
\end{equation}
The expansion coefficient in Equation~(3),
\begin{equation}
C_\ell=\frac{1}{2\ell+1}\sum_m |a_{\ell m}|^2\;,
\end{equation}
is known as the angular power of multipole $\ell$.

\begin{figure}
\vspace{1cm}
\centering
\includegraphics[width=8cm]{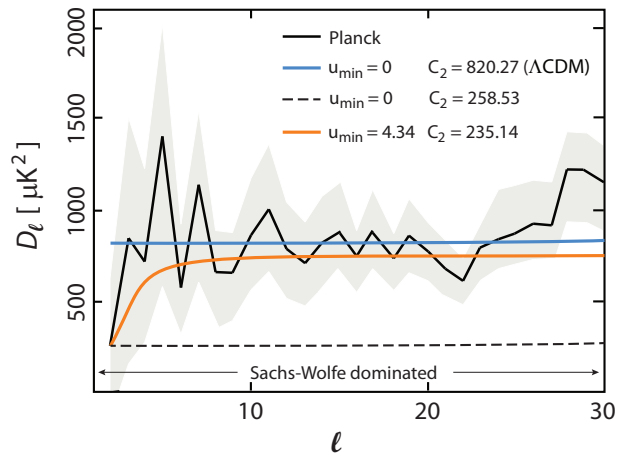}
\caption{Power spectrum (black) estimated with the {\it NLIC} method (Planck 2014a), 
with $1\sigma$ Fisher errors (grey shaded). The {\it Planck} $\Lambda$CDM best fit model,
including solely the Sachs-Wolfe (SW) effect, is shown in blue. The red curve for
$u_{\rm min}\not=0$, and the dashed curve for $u_{\rm min}=0$, also based solely
on SW, are optimized based on the best fit to $C(\cos\theta)$ shown in figure~2. 
SW dominates at large angles $\gg 1-2^\circ$, corresponding to $\ell\lesssim 30$, 
while local physical effects, e.g., BAO, dominate for $\ell\gtrsim 30$.}
\end{figure}

In the ideal case of full-sky coverage, $C(\cos\theta )$ provides a complementary 
means of analyzing CMB
observations instead of the angular power spectrum $C_\ell $. In principle,
$C(\cos\theta )$ contains the same information as the angular power spectrum,
but it provides an easier understanding of the anisotropic structures and
may serve as a complementary means of spherical-harmonic analysis, the most
commonly used method. Some authors
(Smoot et al. 1992; Hinshaw et al. 1996; Kashlinsky et al. 2001; Copi et al. 2015;
L\'opez-Corredoira \& Gabrielli 2013)
have already attempted a direct determination of the anisotropies
correlation function directly in angular space.

Several physical influences contribute to $C_\ell$, some preferentially at
large angles (i.e., $\theta\gg 1-2^\circ$), others---such as baryon acoustic
oscillations (BAO)---predominantly on smaller scales (Meiksin et al. 1999;
Seo \& Eisenstein 2005; Jeong \& Komatsu 2006; Crocce \& Scoccimarro 2006;
Eisenstein et al. 2007; Padmanabhan \& White 2009). 
The 2013 release of the {\it Planck} temperature power
spectrum (Planck 2014a) up to $\ell=30$ is shown in figure~1,
along with two theoretical fits that we shall discuss shortly. In this figure,
$D_\ell\equiv \ell(\ell+1)C_\ell/2\pi$. The well-known dichotomy between
effects at large and small angles allows us to greatly simplify the calculation
of $C_\ell$ for the purpose of this paper. At large angles, corresponding
to $\ell\lesssim 30$, the dominant physical process producing the anisotropies
is the Sachs-Wolfe (SW) effect (Sachs \& Wolfe 1967), representing metric perturbations
associated with scalar fluctuations in the matter field. This effect relates the
anisotropies observed in the temperature today to inhomogeneities of the metric
fluctuation amplitude on the surface of last scattering. For the power-law spectrum
of perturbations in Equation~(2),
and assuming only SW, the angular power (Equation~6) reduces to the integral
expression (Bond \& Efstathiou 1984; Hu \& Sugiyama 1995)
\begin{equation}
C_\ell=N\int_{k_{\rm min}}^\infty k^{n-2}\,j_\ell^2(kc\Delta\tau_{\rm dec})\,dk\;,
\end{equation}
where $j_\ell$ is the spherical Bessel function of order $\ell$, and
$c\Delta\tau_{\rm dec}$ is the comoving radius of the last scattering
surface written in terms of the conformal time difference between
$t_0$ and $t_{\rm dec}$. The normalization constant $N$ is typically determined
by optimizing the fit to the data in figure~1, though in this paper we
will also show the impact of optimizing $N$ for $C(\cos\theta)$. We stress that
the key difference between the conventional approach and the novel idea we
are introducing here, is the appearance of a non-zero lower limit, $k_{\rm min}$,
to the integral in Equation~(7). As we shall see shortly, it is this $k_{\rm min}$
that accounts for the absence of CMB angular correlations at angles $\gtrsim 60^\circ$,
representing a characteristic spatial scale apparently equal to $2\pi$ times the
gravitational radius $R_{\rm h}(t_{\rm dec})=c/H(t_{\rm dec})$ at decoupling.

\begin{figure}
\vspace{1cm}
\centering
\includegraphics[width=8cm]{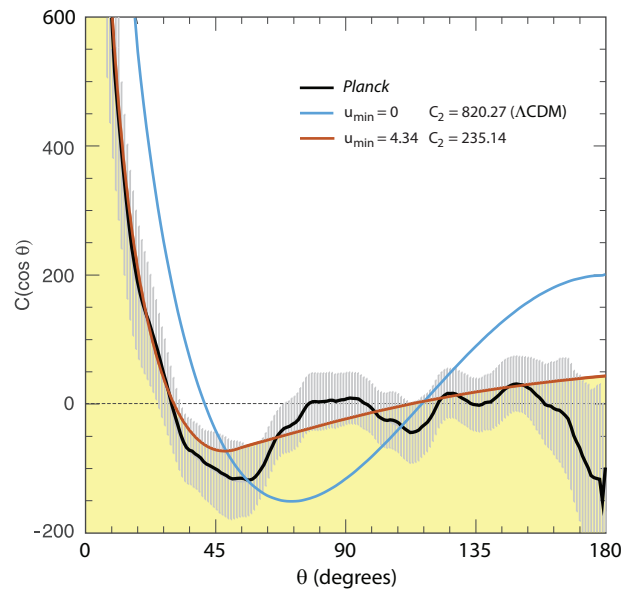}
\caption{Angular correlation function measured with {\it Planck}
(dark solid curve) (Planck 2014a), and associated $1\sigma$ errors
(grey), compared with (blue) the prediction of the conventional inflationary
$\Lambda$CDM, in which $u_{\rm min}=0$ and $C_2$ is optimized using the
power spectrum in figure~1, and (red) truncated inflation, or a non-inflationary
cosmology, with an optimized lower limit $u_{\rm min}=4.34$ and $C_2=235.14$.}
\end{figure}

From the WMAP and {\it Planck} observations, we infer that the power spectrum
in Equation~(2) is very nearly scale free with $n\approx 1$ (Planck 2014a).
Therefore, selecting this value in Equation~(7), and defining the variable
\begin{equation}
u_{\rm min}\equiv k_{\rm min}\,c\Delta\tau_{\rm dec}\;,
\end{equation}
we may recast the expression for $C_\ell$ in the form
\begin{equation}
C_\ell=N\int_{u_{\rm min}}^\infty \frac{j_\ell^2(u)}{u}\,du\;.
\end{equation}
The constant $u_{\rm min}$ is $2\pi$ times
the number of proper maximum wavelengths $\lambda_{\rm max}$ (corresponding
to $k_{\rm min}$) in the proper distance to the last scattering surface
at $t_{\rm dec}$. Its value allows us to determine the angular size $\theta_{\rm max}$
of the largest fluctuation on this surface, using the expression
\begin{equation}
u_{\rm min}=\frac{2\pi}{\lambda_{\rm max}}\,a(t_{\rm dec})\,c\Delta\tau_{\rm dec}\;,
\end{equation}
where $a(t)$ is the expansion factor. So using the standard definition of
the angular-diameter distance, this simply reduces to the form
$\theta_{\rm max}={2\pi/ u_{\rm min}}$.

\section{Discussion}
The CMB angular correlation function $C(\theta)$ based on the 
{\it Planck} 2013 release (Planck 2014a) has been calculated
by averaging all temperature pairs in the sky separated by angles
$\theta$ inside bins of size $1^\circ$. We used the component-separated
CMB map resulting from the Needlet Internal Linear Combination (NILC)
algorithm, downloaded from the Planck Legacy
Archive (PLA\footnote{http://www.cosmos.esa.int/web/planck/pla}). This
map was smoothed to an angular resolution of $1^\circ$ and degraded to
Healpix $N_{\rm side}=64$ (pixel size $0.92^\circ$). In order to avoid
foreground contamination from residual Galactic emission or point
sources, we excluded from our analysis all pixels affected by the
multiplication of the Commander, SEVEM and SMICA masks, which keeps $67\%$
of the sky. (Note that we did not take into account the NILC mask
because it removes a significantly smaller fraction of the sky.) The
$1\sigma$ error bars were computed through a comparison of the
angular correlation function calculated for 200 randomly-rotated
maps based on the original distribution.

The calculated angular power $C_\ell$ for multipoles $2\le \ell\le 30$ is shown
for the conventional $\Lambda$CDM (blue curve) in figure~1, optimized to fit
the {\it Planck} 2013 release. Also, for comparison, we show in this
figure a fit for $u_{\rm min}=0$ (dashed) optimized using $C(\cos\theta)$ in
figure~3 instead of the angular-power spectrum. In figure~1, the theoretical curves
take into account only SW, ignoring
the physical influences (such as BAO) that would dominate on small angular
scales (i.e., at $\ell\gtrsim 30$). It is well known that the latter depend, at
most, only weakly on the cosmology, and are therefore degenerate among different
models (Scott et al. 1995).

The fit for $u_{\rm min}\not=0$ is optimized using $C(\cos\theta)$ in figure~2.
For the $\Lambda$CDM (blue) curve, we have followed the conventional
approach of first fitting the temperature spectrum to determine the
normalization constant $N$ in Equation~(9), which is then used together
with Equation~(3) to produce the angular correlation function.
It has been known since the early WMAP release (Spergel et al. 2003)
that the $\Lambda$CDM curve is inconsistent with the measured
$C(\cos\theta)$, but when viewed here in comparison with the curve
corresponding to $u_{\rm min}\not=0$, its lack of adequate confirmation
by the data is even more glaring. The red curve in figure~2 shows the
best fit attainable with $u_{\rm min}\not=0$. Based on the 
{\it Planck} 2013 data release, we find an optimized
value $u_{\rm min}=4.34\pm 0.50$, corresponding to a maximum fluctuation
size $\theta_{\rm max}\approx 83^\circ$ in the plane of the sky.
For comparison with the value of $k_c$ estimated earlier by
Niarchou et al. (2004) using WMAP data, we determine for $\Lambda$CDM
(in which $z_{\rm cmb}=1080$) that this measurement of $u_{\rm min}$ corresponds
to a maximum fluctuation size $\lambda_{\rm max}\sim 20$ Mpc at decoupling,
about twice the value corresponding to their characteristic wavenumber
$k_c$. Of course, with the benefit of using the more precise 
{\it Planck} data, and our focus on the large-angle anomalies rather than
the entire CMB power spectrum, we also conclude that a cutoff in $P(k)$ is 
favoured at a much higher level of significance than before, now in excess 
of $\sim 8\sigma$.

This $\sigma=0.5$ error in the measurement of $u_{\rm min}$ was obtained from 
a Monte Carlo analysis sampling the
variation of $C(\cos\theta)$ within the measurement errors. The $C(\cos\theta)$
points are highly correlated, but our analysis circumvents this problem
with the Monte Carlo procedure. First, we generate 100 mock CMBR catalogs
(using standard cosmology with an angular function $C_0(\cos\theta)$,
and we measure the two-point correlation $C_i(\cos\theta)$ in each case.
From these, we obtain $\Delta C_i(\cos\theta)\equiv C_i(\cos\theta)-
C_0(\cos\theta)$, and then calculate $u_{{\rm min},\,i}$ for $C(\cos\theta)=
C_{\it Planck}(\cos\theta)+\Delta C_i(\cos\theta)$ for each realization $i$.
Next, we examine the distribution of $u_{{\rm min},\,i}$ (which is roughly
Gaussian) and determine its average value and the r.m.s., within which one
finds 68 of the 100 values: this yields the quoted result $u_{\rm min}=4.34\pm 
0.50$. This is not equivalent to a simple $\chi^2$ fitting with the correlated
error bars shown in figures~2 and 3. Such a simple $\chi^2$ procedure
would instead have given $u_{\rm min}=4.34^{+0.10}_{-0.12}$, i.e., a much
smaller error for $u_{\rm min}$. This difference stems precisely from the
fact that the $C(\cos\theta)$ points are correlated. Therefore the $68\%$
confidence limit for this value of $u_{\rm min}$ suggests that a
theoretical best fit to the measured angular correlation function with
$u_{\rm min}=0$ is rejected at $8.6\,\sigma$.

\begin{figure}
\vspace{1cm}
\centering
\includegraphics[width=8cm]{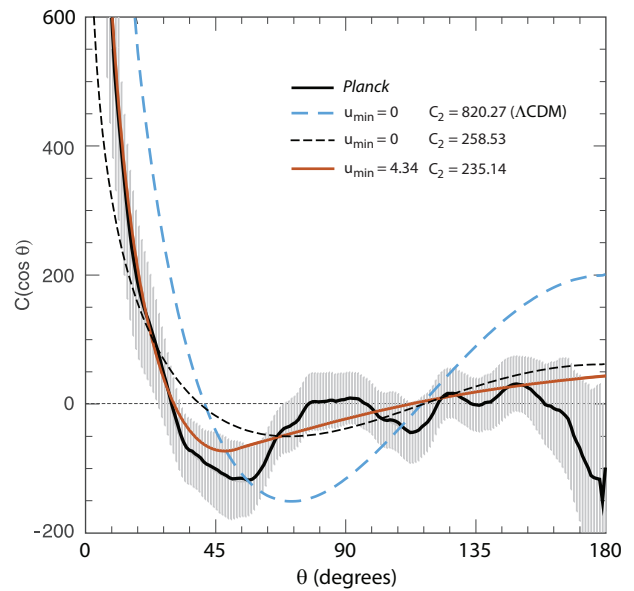}
\caption{Impact of $u_{\rm min}$ on the angular correlation function with 
truncated inflation, or a non-inflationary cosmology. The conventional 
$\Lambda$CDM curve corresponds to $u_{\rm min}=0$, i.e., an unconstrained 
inflationary power spectrum. The Planck measurement is shown as a black 
solid curve.}
\end{figure}

From figure~2 we see that there are several good reasons for preferring the
$u_{\rm min}=4.34$ fit over a model with $u_{\rm min}=0$: (1) {\it Planck}
has confirmed that $C(\cos\theta)\approx 0$ at angles $\theta\gtrsim 60^\circ
-70^\circ$, in sharp contrast to the prediction of the conventional inflationary
$\Lambda$CDM; (2) the model with $u_{\rm min}=4.34$ correctly reproduces the
minimum of $C(\cos\theta)$ at $\approx 45^\circ-50^\circ$, and (3) it actually
fits the measured curve beyond $\sim 50^\circ$ quite well, fully consistent
with the measurement errors.

Suppose we were to optimize $C_2$ with $u_{\rm min}=0$ based on the angular
correlation data in figure~2, instead of the power spectrum in figure~1.
This third case is shown as a dashed black curve in figure~3. Certainly,
the fit to $C(\cos\theta)$ has improved, though there is still significant
tension with theory at $\theta\lesssim 20^\circ$, and the best-fit curve
completely misses the minimum of $C(\cos\theta)$ at $\theta\sim 50^\circ$.
In addition, with this optimized value of $C_2$ (i.e., $258.53\;\mu$K$^2$),
the theoretical angular power spectrum is now a very poor fit to the measured
$D_\ell$'s shown in figure~1. By far, the best fit to the angular power
spectrum and correlation function is realized when $u_{\rm min}=4.34\pm0.50$.

\section{Conclusion}
The $S_{1/2}$ statistic (Spergel et al. 2003) has traditionally been used to
categorize the degree to which the measured angular correlation at large
angles is deficient compared to theoretical predictions. This quantity
is basically an integral of $C(\theta)^2$ over angles, from $\cos(\theta)=-1$
to $\cos(\theta)=1/2$. With Monte-Carlo simulations, one can build a
distribution of $S_{1/2}$ values, thereby assigning a probability that
the observed angular correlation function could be drawn randomly as a
result of cosmic variance from the predicted function. The $p$-values
quoted earlier in this paper for conventional $\Lambda$CDM are
estimated using this comparison.

Our analysis in this paper is superior to $S_{1/2}$ for several
reasons. First, $S_{1/2}$ represents an integrated quantity, from $\theta\sim 
60^\circ$ to $180^\circ$, designed to find a deficiency in signal. Second,
it completely ignores the comparison between theory and observation at angles
$\lesssim 60^\circ$, where the tension can be just a large as it is at
$\theta\gtrsim 60^\circ$. Both of these limitations with $S_{1/2}$ make
it an inferior statistic to use in this work compared to our approach
of actually fitting the $C(\theta)$ data by optimizing the value
of $u_{\rm min}$. The fact that the conventional inflationary paradigm
is disfavoured by these data in comparison to a model with $k_{\rm min}>0$
is supported by both approaches. But the optimization carried out in this
paper goes significantly beyond this level by demonstrating that
$k_{\rm min}=0$ is ruled out at a confidence level of $\sim 8\sigma$.

A $k>k_{\rm min}$ constraint on $P(k)$ is inconsistent with purely slow-roll
inflationary cosmology, which instead predicts that fluctuations would have
exited and re-entered the horizon prior to decoupling, resulting in $k_{\rm min}=0$
and strong correlations at all angles. On the other hand, such a result is fully
consistent with the predictions of the $R_{\rm h}=ct$ Universe (Melia 2007,2013;
Melia \& Shevchuk 2012), in which fluctuations might have
emerged near the Planck scale, equal to the Hubble radius $R_{\rm h}$ at the
Planck time. The $k_{\rm min}$ might therefore correspond
to the horizon size at the surface of last scattering, since only fluctuations
with $\lambda_{\rm max}\sim 2\pi R_{\rm h}$ would have continued to grow
towards $t_{\rm dec}$.

The fact that the measured CMB angular correlation function strongly
favors a modification to conventional inflation, or eliminating it all
together, is an important validation of other recent results showing
similar trends (Wei et al. 2016).  Looking forward to the next
generation of observations and simulations, the introduction of a
non-zero $k_{\rm min}$ should be quite straightforward to implement.
The introduction of a $k_{\rm min}$ will not affect many other kinds
of observation. The optimized value we have found here corresponds to a
maximum fluctuation angle in the sky of about $83^\circ$. As such, this will
have no impact on the optimization of the power spectrum, since $k_{\rm min}$
affects only the far-left (i.e., $\ell\sim 1-10$) portion of $D_\ell$ in
figure~1. By comparison, the first acoustic peak in the power spectrum is
centered at $\ell\sim 200$ corresponding to an angle of $\sim 1^\circ$.
For similar reasons, such a large maximum fluctuation angle is unlikely to
affect other measurements. At least for now, the focus with $k_{\rm min}$
should be on refining the measurement of the angular correlation function
of the CMB.

From a simulational point of view, however, we have so far ignored the
integrated Sachs Wolfe (ISW) effect, which arises due to the passage of
light from the surface of last scattering to our location (Bond \& Efstathiou 1984).
The ISW is not negligible, though the early-time contribution is typically 
combined with SW at last scattering, and is incorporated into Equation~(7). The 
late-time ISW is smaller, but nonetheless present, so a complete accounting of 
the impact of $k_{\rm min}$ will eventually need to include its influence. We do
not expect our results to change substantially.

\begin{acknowledgements}
We are very grateful to Ricardo G\'enova-Santos for help calculating 
the angular correlation function using the {\it Planck} 2013 data
release. FM acknowledges partial support from the Chinese State Administration of 
Foreign Experts Affairs under grant GDJ20120491013. MLC was supported by 
grant AYA2015-66506-P of the Spanish Ministry of Economy and 
Competitiveness.
\end{acknowledgements}

\end{document}